\documentclass[conference]{IEEEtran}
\IEEEoverridecommandlockouts
\usepackage{cite}
\usepackage{amsmath,amssymb,amsfonts}
\usepackage{graphicx}
\usepackage{textcomp}
\usepackage{xcolor}
\usepackage{booktabs}
\usepackage{url}
\def\BibTeX{{\rm B\kern-.05em{\sc i\kern-.025em b}\kern-.08em T\kern-.1667em\lower.7ex\hbox{E}\kern-.125emX}}

\usepackage{fancyhdr}
\fancypagestyle{firstpage}{%
 \fancyhf{}%
 \fancyfoot[L]{\normalsize\textbf{979-8-3195-2149-1/26/\$31.00~\copyright2026 IEEE}}%
}

\begin{document}

\title{Pre-Whitening and BCJR Posterior Distillation for Bi-LSTM Detection in Faster-than-Nyquist Signaling}

\author{\IEEEauthorblockN{Nurettin Safak, Osman Tokluoglu, Enver Cavus}
\IEEEauthorblockA{Department of Electrical and Electronics Engineering,
Ankara Yildirim Beyazit University, Ankara, Turkiye\\
nsafaked@gmail.com, otokluoglu@aybu.edu.tr, ecavus@aybu.edu.tr}}

\maketitle
\thispagestyle{firstpage}   

\begin{abstract}
Recurrent detectors such as bidirectional long short-term memory (Bi-LSTM)
networks are low-complexity alternatives to the optimal
Bahl-Cocke-Jelinek-Raviv (BCJR) detector for faster-than-Nyquist (FTN)
signaling. Motivated by convolutional detectors that build the intersymbol
interference (ISI) structure into their architecture, we ask whether processing
nested ISI windows in separate recurrent branches improves the bit error rate
(BER) of a Bi-LSTM. Across roughly 260 controlled trainings it does not: at a
matched parameter budget and a matched readout, the multi-window architecture
never significantly beats a plain Bi-LSTM. Nested windowing is an invertible
rearrangement that adds no information, extra branches only add bottlenecks, and
a distillation diagnostic shows the network is already near optimal for its
window. The limitation is therefore the observation model, not the architecture.
Keeping the architecture fixed, we pre-whiten the input, restoring the
conditional independence that colored matched-filter noise violates, and distill
the BCJR soft posterior into the network. With 3.4\% more parameters this reaches
1.05 times the BCJR BER at a compression factor of 0.8 and 1.89 times at 0.7,
improving to 1.47 times when the whitened window is widened. The 23.7\% BER
reduction at 0.8 requires an ill-conditioned ISI matrix but is not monotone in
the conditioning, and it holds across five independent noise realizations and a
symbol-level McNemar test.
\end{abstract}


\section{Introduction}
Faster-than-Nyquist (FTN) signaling improves spectral efficiency by
transmitting symbols faster than the Nyquist rate, at the cost of deliberate
intersymbol interference (ISI)~\cite{mazo,anderson}. The optimal detector is the
Bahl--Cocke--Jelinek--Raviv (BCJR) maximum a posteriori (MAP)
algorithm~\cite{bcjr}, whose complexity grows exponentially with the channel
memory. Reduced-complexity variants such as M-BCJR~\cite{mbcjr} lower this cost
but remain model-based, motivating low-complexity learned detectors that trade a
small bit-error-rate (BER) penalty for a large complexity saving. Recurrent networks, in particular long short-term
memory (LSTM) and bidirectional LSTM (Bi-LSTM) variants, and more recently
convolutional and self-attention Transformer networks have been proposed as
standalone FTN detectors~\cite{gru,bigru,cnnfk,defilippo,paul,transformerftn,elmanrnn,kanmlp}.

A recent CNN detector~\cite{cnnfk} obtains near-BCJR performance for
$\tau\ge0.7$ by \emph{structuring} the network: fixed, distance-indexed kernels
with domain-informed masking process the center symbol together with its
$\pm i$ neighbors, one layer per distance. This raises a natural question for
recurrent detectors: \emph{does transferring the same ISI-aware, multi-window
structure to a Bi-LSTM improve its BER?}

That CNN gain, however, mainly compensates the translation-equivariance
(``position-blindness'') of convolution~\cite{cnnfk,domainaware}---an issue a
recurrent network does not have, which is precisely why the multi-window
transfer fails here. Closest to
our setting, a position-aware, attention-enhanced Bi-GRU detector has recently
been reported for FTN~\cite{bigru}; that making a \emph{recurrent} detector
position-aware pays off is consistent with our finding that the readout position
alone accounts for a $22$--$33\%$ BER change (Section~\ref{sec:method}), although
here that effect is obtained without any attention mechanism, and it is
orthogonal to the observation-model fix we propose. Neural FTN
equalizers using a super-minimum-phase (Forney) transform have been
reported~\cite{paul}, but without an ablation isolating the value of whitening
and without distillation, and only for $\tau\le0.5$; a recent CNN FTN
detector~\cite{defilippo} explicitly \emph{chooses not to} whiten. Model-based
detectors that unfold the sum-product algorithm into a network have also been
proposed for FTN~\cite{sumproduct}, and BCJRNet/ViterbiNet~\cite{bcjrnet,viterbinet} \emph{learn} the
BCJR computation rather than distilling its posterior into a plain detector, and
knowledge distillation in optical equalization has been NN-to-NN
(biGRU$\to$CNN) rather than from the optimal detector~\cite{opticalkd}. To our
knowledge, BCJR-posterior distillation into a neural FTN detector, and an
ablation isolating whitening for a recurrent FTN detector, are both new.

We answer this question negatively and, more importantly, explain why. We then
show that the real bottleneck is the observation model, not the architecture,
and propose a fix that leaves the architecture untouched. Our contributions are:
\begin{itemize}
\item A fair, readout-matched study ($\approx 260$ trainings) showing that
nested multi-window / distance-isolated recurrent branches do not beat a plain
Bi-LSTM under a matched parameter budget (Section~\ref{sec:method}).
\item A distillation \emph{diagnostic} proving that the plain network is already
near window-optimal, localizing the gap to the observation model
(Section~\ref{subsec:diag}).
\item An observation-model-aware detector: pre-whitening plus BCJR soft-posterior
distillation, architecture unchanged, that closes most of the BER gap
(Sections~\ref{sec:method} and~\ref{sec:res}); to our knowledge the first
BCJR-posterior distillation for FTN and the first ablation isolating the value
of whitening for a recurrent FTN detector.
\end{itemize}

\section{System Model}\label{sec:sys}
\begin{figure}[t]
\centering
\includegraphics[width=0.99\columnwidth]{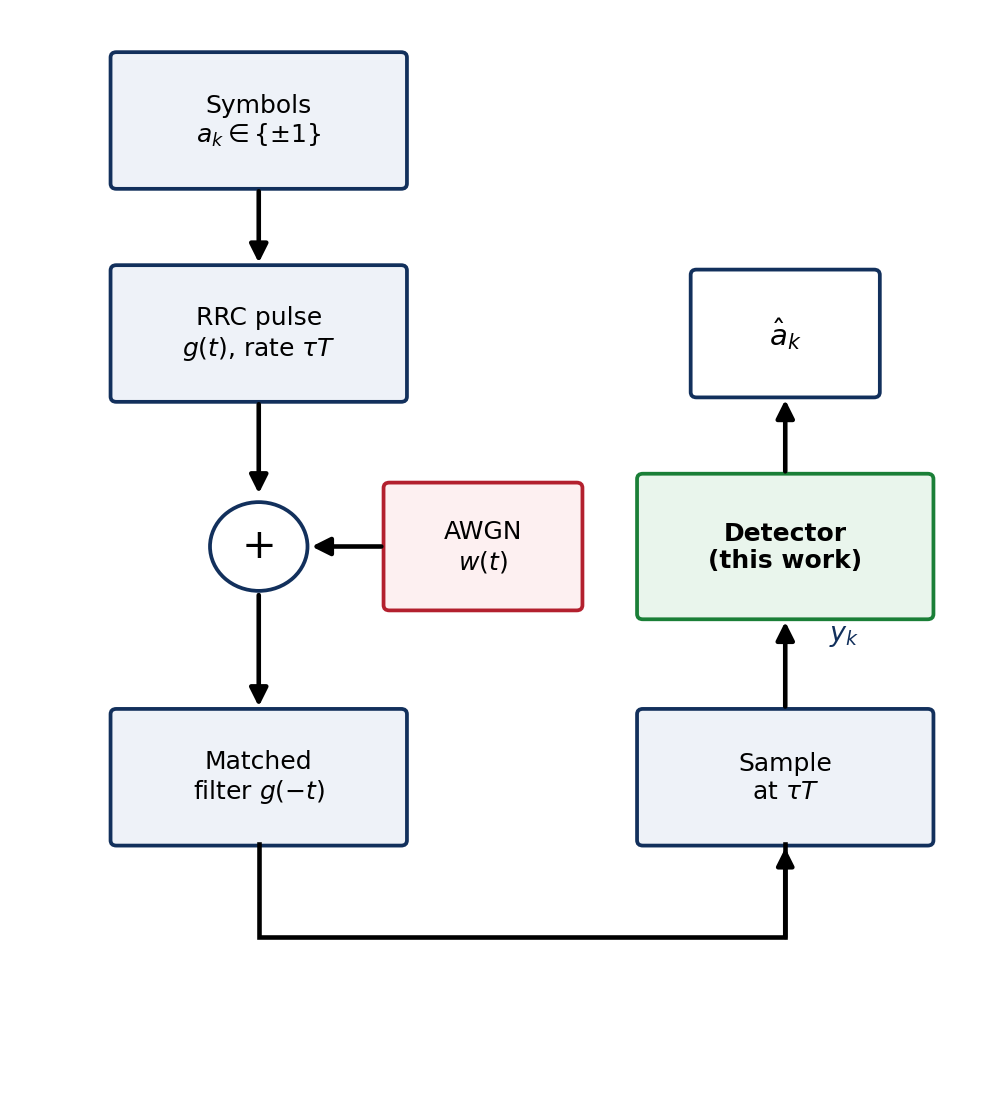}
\caption{FTN system model. Symbols are shaped by an RRC pulse at the faster rate
$\tau T$, corrupted by AWGN, and matched-filtered; the sampled output $y_k$
follows the Ungerboeck model with \emph{colored} noise whose correlation is the
ISI sequence itself.}
\label{fig:sysmodel}
\end{figure}

With a unit-energy root-raised-cosine (RRC) pulse of roll-off $\beta$ and
compression factor $\tau\in(0,1)$, the matched-filter output at symbol index
$k$ under the Ungerboeck observation model~\cite{ungerboeck} is
\begin{equation}
y_k = \sum_{i=-N}^{N} x_i\, a_{k-i} + w_k,
\label{eq:model}
\end{equation}
where $a_k\in\{\pm1\}$ are BPSK symbols, $x_i$ are the (known) ISI coefficients
obtained from the pulse autocorrelation, and $N$ is the one-sided ISI length.
Crucially, the noise is \emph{colored}: $\mathbb{E}[w_k w_l]=(N_0/2)\,x_{k-l}$,
i.e.\ its correlation is the ISI sequence itself. The exact log-likelihood
$\ln p(\mathbf{y}\mid\mathbf{a})\propto 2\mathbf{a}^\top\mathbf{y}-\mathbf{a}^\top
\mathbf{X}\mathbf{a}$ with $\mathbf{X}$ the Toeplitz ISI matrix is a sufficient
statistic only over the \emph{whole} sequence; the BCJR algorithm evaluates the
corresponding MAP posterior. All learned detectors below observe a length-$2N{+}1$
window $\mathbf{y}_k=[y_{k-N},\dots,y_k,\dots,y_{k+N}]$ (Fig.~\ref{fig:sysmodel}) and estimate the center
symbol $a_k$; we report BER normalized by the BCJR (MAP) BER, so that
$1.00\times$ is optimal.

\section{Observation-Model-Aware Bi-LSTM Detection}\label{sec:method}

\subsection{Multi-window architectures}
The plain baseline feeds $\mathbf{y}_k$ into a single Bi-LSTM. The multi-window
detector, following the specification inspired by~\cite{cnnfk}, builds nested
windows $W_m(k)=[y_{k-m},\dots,y_{k+m}]$, $m=1,\dots,N$, and processes each in a
\emph{separate} Bi-LSTM branch; the branch outputs are concatenated and passed
through a shared dense head. A distance-isolated variant instead feeds each
branch only the symmetric triplet $[y_{k-i},y_k,y_{k+i}]$, mirroring the CNN
kernel of~\cite{cnnfk}.

\subsection{Readout is a confound, and it dominates}
\begin{figure}[t]
\centering
\includegraphics[width=0.99\columnwidth]{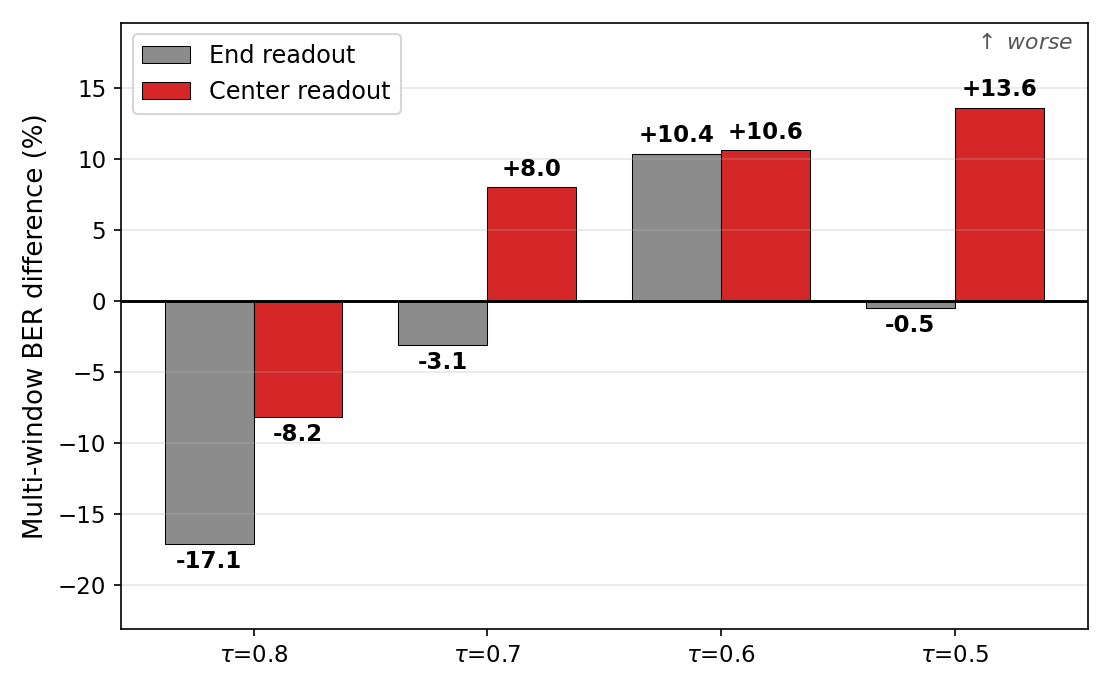}
\caption{Architecture effect at $8$\,dB (readout held fixed): multi-window minus
plain Bi-LSTM BER, in percent, for each compression factor. Once the readout is
matched, the multi-window architecture gives no robust advantage---it wins only
at $\tau=0.8$ and loses at $\tau=0.7,0.6,0.5$.}
\label{fig:arch}
\end{figure}

\begin{figure}[t]
\centering
\includegraphics[width=0.99\columnwidth]{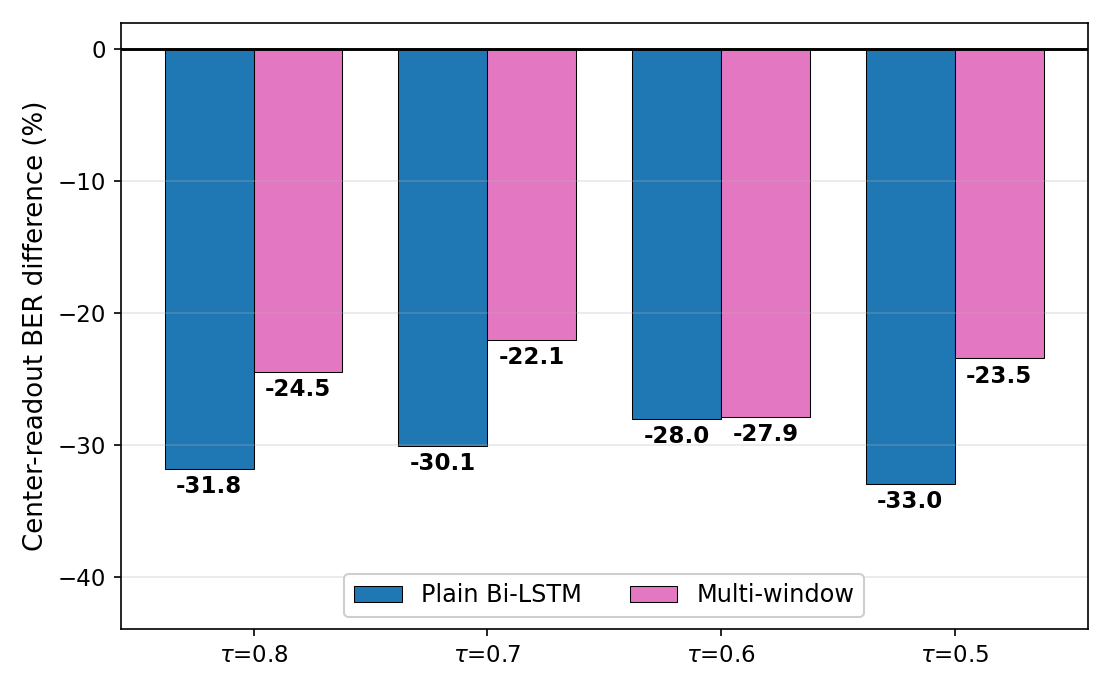}
\caption{Readout effect at $8$\,dB (architecture held fixed): center minus end
readout BER, in percent, for both the plain and the multi-window model. The
change is large and consistent ($8/8$ significant, $22$--$33\%$), dominating the
architecture effect.}
\label{fig:readout}
\end{figure}

Bi-LSTM detectors admit two readouts: the concatenated \emph{end} states
$[\mathbf{h}^{\rightarrow}_{\text{last}},\mathbf{h}^{\leftarrow}_{\text{last}}]$,
or the \emph{center} hidden state $\mathbf{o}[N]$. Under end readout the target
symbol $a_k$ is the \emph{oldest} input for both directions and is diluted
through $N$ additional gated updates; the multi-window branches, being shorter,
implicitly read closer to the center. To separate architecture from readout we
run a $2\times2$ design (architecture $\times$ readout) at nearly equal
parameters ($5809$ plain vs.\ $5625$ multi-window, a $3.2\%$ difference in
favour of the plain model) and identical protocol. Table~\ref{tab:decisive} and
Figs.~\ref{fig:arch} and~\ref{fig:readout} show that \emph{readout} accounts for a $22$--$33\%$ BER
change (significant in $8/8$ cases), whereas the multi-window architecture, once
readout is matched, beats the plain model only at $\tau=0.8$ and loses at
$\tau=0.7,0.6,0.5$.

\begin{table}[t]
\caption{$2\times2$ decisive experiment: BER normalized by MAP at $8$\,dB,
$5$ seeds, matched parameters ($5809$ vs.\ $5625$), identical protocol.}
\label{tab:decisive}
\centering
\begin{tabular}{lcccc}
\toprule
Model & $\tau{=}0.8$ & $\tau{=}0.7$ & $\tau{=}0.6$ & $\tau{=}0.5$\\
\midrule
Plain, end readout        & 1.95 & 3.41 & 2.42 & 2.95\\
Plain, center readout     & \textbf{1.33} & \textbf{2.39} & \textbf{1.74} & \textbf{1.98}\\
Multi-window, end readout & 1.62 & 3.31 & 2.67 & 2.93\\
Multi-window, center      & 1.22 & 2.58 & 1.93 & 2.25\\
\bottomrule
\end{tabular}
\end{table}

\subsection{Why the architecture cannot help}
Two structural arguments explain the negative result. First, nested windowing is
\emph{information-preserving}: since $W_N=\mathbf{y}_k$, the map
$\mathbf{y}_k\mapsto(W_1,\dots,W_N)$ is injective, hence
$I(W_1,\dots,W_N;a_k)=I(\mathbf{y}_k;a_k)$. The preprocessing adds no
information; any gain must come from inductive bias alone. Second, each branch
irreversibly compresses its window \emph{before} fusion; with $N$ branches the
Markov chain $a_k\!\to\!\mathbf{y}_k\!\to\!(\mathbf{h}_1,\dots,\mathbf{h}_N)\!\to\!\hat a_k$
has $N$ bottlenecks instead of one, and by the data-processing inequality this
can only reduce $I(\hat a_k;a_k)$. The distance-isolated variant is worse: a
branch seeing only $[y_{k-i},y_k,y_{k+i}]$ cannot separate its own
distance's contribution to $y_k$ from that of all other distances, which acts as
self-noise; this collapses precisely at low $\tau$ ($8.96\times$ vs.\
$3.61\times$ MAP at $\tau=0.7$, i.e.\ $2.48\times$ worse than the plain
baseline trained in the same run under the same protocol).

\subsection{Distillation diagnostic: already window-optimal}\label{subsec:diag}
To locate the gap we exploit a property of soft-label
distillation~\cite{hinton}: because the
minimizer of the soft cross-entropy against the full-BCJR posterior is, by the
tower rule, exactly the window-conditional posterior
$\mathbb{E}[\,\mathbb{E}[a_k\mid\mathbf{y}_{\text{full}}]\mid W]=\mathbb{E}[a_k\mid W]$,
distillation is \emph{unbiased} and only helps if a learning gap exists.
Distilling the BCJR posterior into the plain $y$-input network yields
\emph{no} significant gain (Table~\ref{tab:main}, ``$+$distill''). Hence the
network is already near window-optimal in the $y$ domain, and no
architectural change can help---the gap is in the observation model.

\begin{figure}[t]
\centering
\includegraphics[width=0.99\columnwidth]{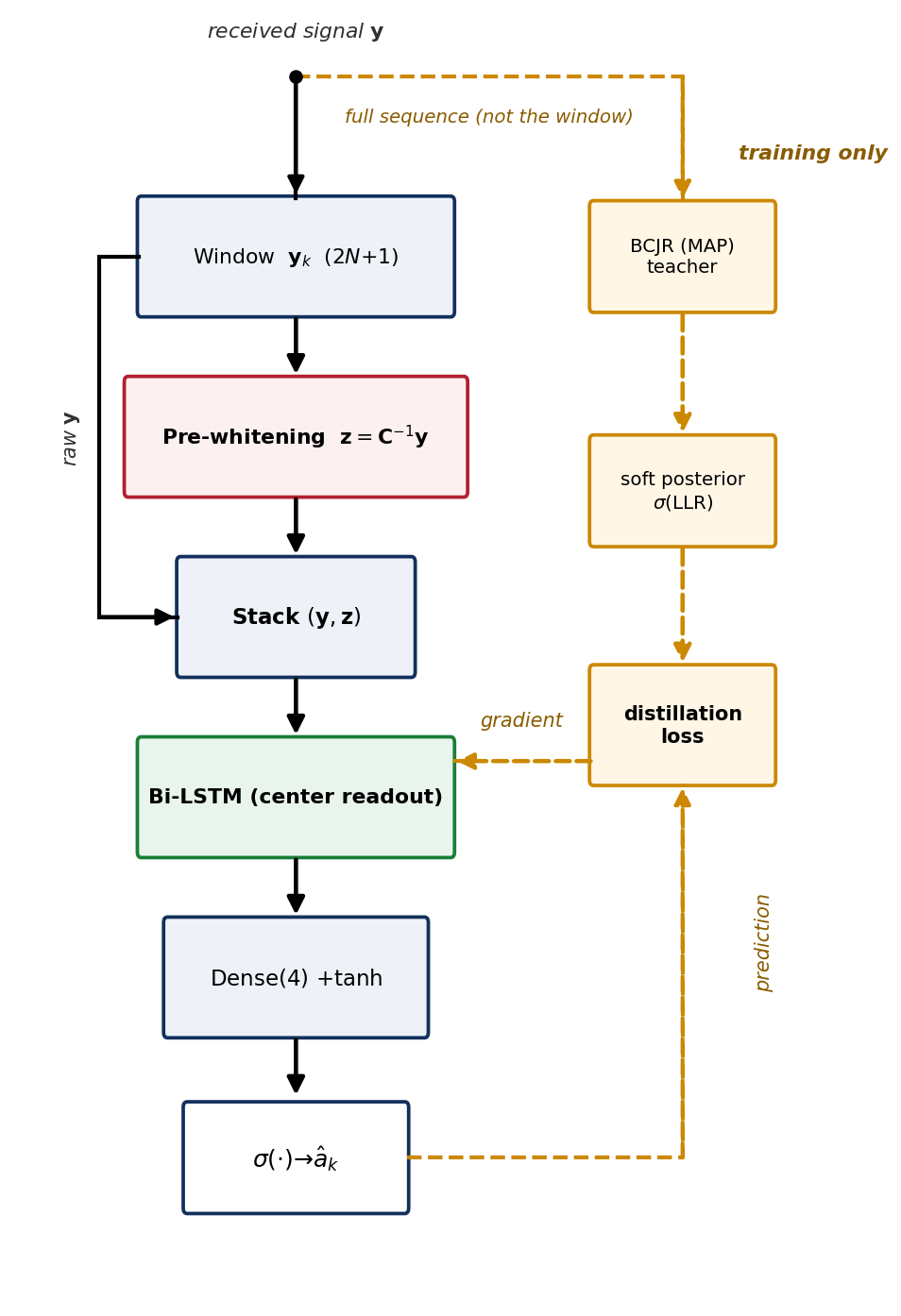}
\caption{Proposed observation-model-aware detector. The architecture (a plain
Bi-LSTM with center readout) is unchanged; only the input representation
(adding the whitened channel $\mathbf{z}=\mathbf{C}^{-1}\mathbf{y}$) and the
training target (BCJR soft posterior) change, at a cost of $+3.4\%$ parameters.}
\label{fig:pipeline}
\end{figure}

Fig.~\ref{fig:pipeline} summarizes the proposed detector: the same Bi-LSTM
receives the raw window together with its whitened version, and is trained
against the soft posterior of the optimal BCJR detector; the teacher is used
only during training.

\subsection{Pre-whitening}
A Bi-LSTM implicitly assumes conditionally independent observations, which the
colored noise of~\eqref{eq:model} violates. We factor $\mathbf{X}=\mathbf{C}\mathbf{C}^\top$
(banded Cholesky) and define the whitened signal $\mathbf{z}=\mathbf{C}^{-1}\mathbf{y}$,
so that $\mathbf{z}=\mathbf{C}^\top\mathbf{a}+\mathbf{n}$ with white $\mathbf{n}$
(measured residual autocorrelation drops from $0.351$ to $0.038$ at $\tau=0.7$).
Below the folded-spectrum zero the Toeplitz matrix $\mathbf{X}$ loses positive
definiteness numerically, so the Cholesky step is applied to
$\mathbf{X}+\delta\mathbf{I}$ with an adaptive ridge
$\delta=1.35\max(0,-\min_\omega G(\omega))+10^{-3}$, giving
$\delta=10^{-3},10^{-3},2\!\times\!10^{-3},3.0\!\times\!10^{-2},1.03\!\times\!10^{-1}$
for $\tau=0.9,\dots,0.5$. At $\tau=0.5$ this is already $10\%$ of $x_0$, i.e.\
the whitening transform itself becomes inexact---which is the main reason the
method degrades there. Because $\mathbf{C}^{-1}$ is an IIR filter, the whitened window carries
information from \emph{outside} the raw window---this is precisely what
nested rearrangement cannot do, and it lifts the attainable window-MAP BER
(from $1.68\times10^{-3}$ to $6.98\times10^{-4}$ at $\tau=0.7,8$\,dB, $N=6$).
The network receives two channels $(\mathbf{y},\mathbf{z})$; the architecture is
unchanged.

\subsection{BCJR soft-posterior distillation}
At BER $\sim10^{-3}$ the hard-label gradient is informative only at rare error
events. We instead regress the network onto the soft posterior
$\sigma(\mathrm{LLR})$ of the full BCJR detector, run offline on whitened
samples. The teacher is used \emph{only during training}; at inference the
detector is the same small Bi-LSTM. Whitening raises the ceiling; distillation
lets the network reach it. The two are weak alone and strong together---but only
in the regime where both mechanisms apply: this holds at $\tau=0.8,0.7,0.6$,
whereas at the well-conditioned $\tau=0.9$ there is no ceiling to raise
(distillation alone is best) and at $\tau=0.5$ the ridge-regularized whitening is
too inexact for the teacher to help (whitening alone is best); see
Table~\ref{tab:main}.

\section{Simulation Results}\label{sec:res}

\subsection{Experimental setup}
BPSK over AWGN, RRC $\beta=0.35$, $\tau\in\{0.9,0.8,0.7,0.6,0.5\}$. Windows use
$N=6$ (and $N=8$ in the whitened domain). Each of $10^6$ symbols per
$(\tau,\mathrm{SNR})$ is split into a $500$k training half and a disjoint
$500$k test half. Training uses the SNR set $\{7,8,9,10\}$\,dB (pooled), while
evaluation spans $0$--$10$\,dB and all reported ratios are taken at $8$\,dB.
Models are trained with Adam (lr $10^{-3}$, StepLR
$\gamma=0.9$/$5$ epochs), $20$ epochs, batch $1000$, over $5$ seeds; readout is
center-aligned and the dense head ($4$-unit, $\tanh$) is shared across all
compared models. The plain and whitened detectors have $5809$ and $6009$
parameters respectively ($+3.4\%$). An independent Ungerboeck BCJR provides the
MAP reference.

\subsection{Main results}
\begin{figure}[t]
\centering
\includegraphics[width=0.99\columnwidth]{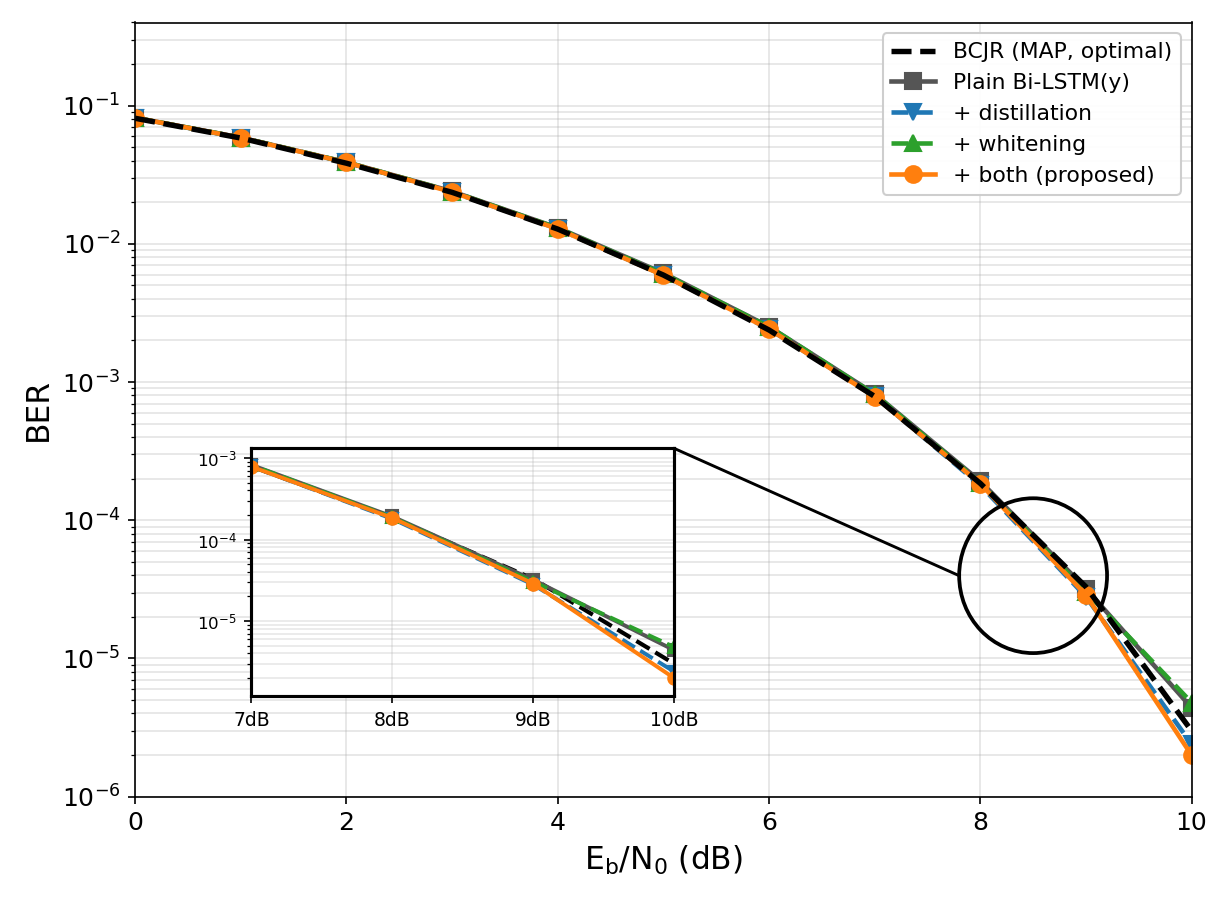}
\caption{BER vs.\ SNR at $\tau=0.9$ ($\beta=0.35$, $\kappa=1.8$; well-conditioned).
All methods coincide with BCJR ($1.04\times\!\to\!0.99\times$ MAP at $8$\,dB);
no room to improve. Inset: $7$--$10$\,dB detail.}
\label{fig:ber09}
\end{figure}

\begin{figure}[t]
\centering
\includegraphics[width=0.99\columnwidth]{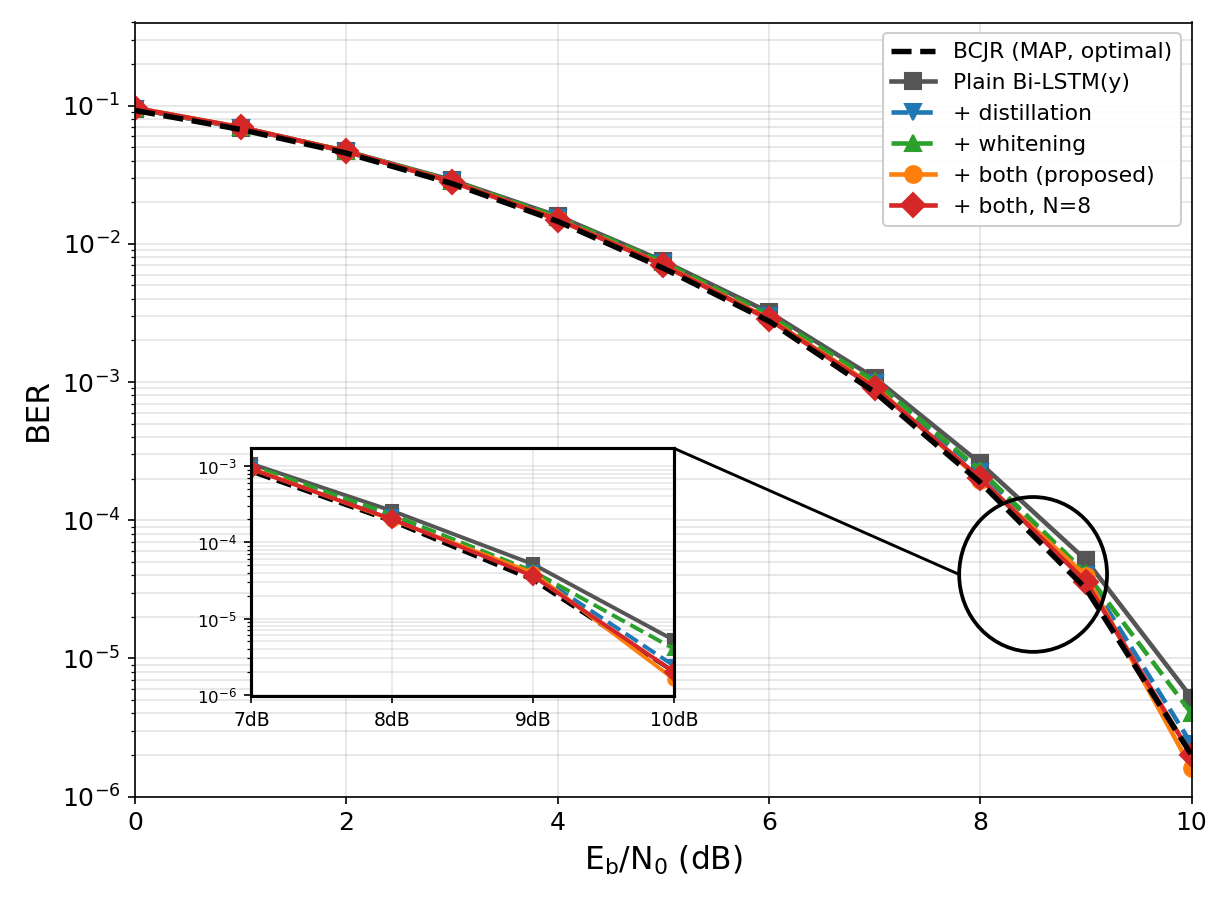}
\caption{BER vs.\ SNR at $\tau=0.8$ ($\beta=0.35$, $\kappa=6.3$). The proposed
detector approaches BCJR ($1.37\times\!\to\!1.05\times$ MAP at $8$\,dB, and
$1.07\times$ with $N{=}8$). Inset: $7$--$10$\,dB detail.}
\label{fig:ber08}
\end{figure}

\begin{figure}[t]
\centering
\includegraphics[width=0.97\columnwidth]{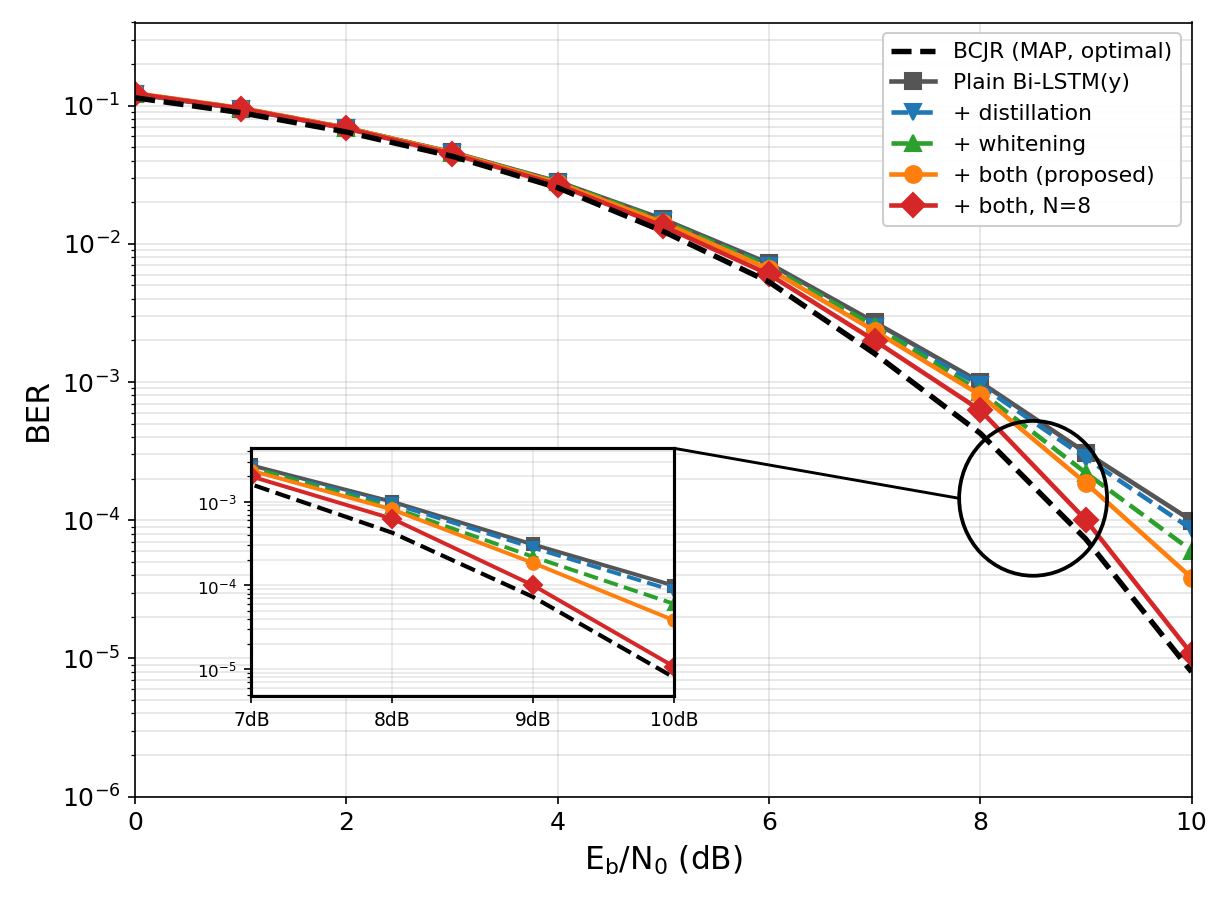}
\caption{BER vs.\ SNR at $\tau=0.7$ ($\beta=0.35$, $\kappa=81$), the largest gap
($2.34\times\!\to\!1.89\times$ MAP at $8$\,dB); $N{=}8$ whitening reaches
$1.47\times$. Inset: $7$--$10$\,dB detail.}
\label{fig:ber07}
\end{figure}

\begin{figure}[t]
\centering
\includegraphics[width=0.97\columnwidth]{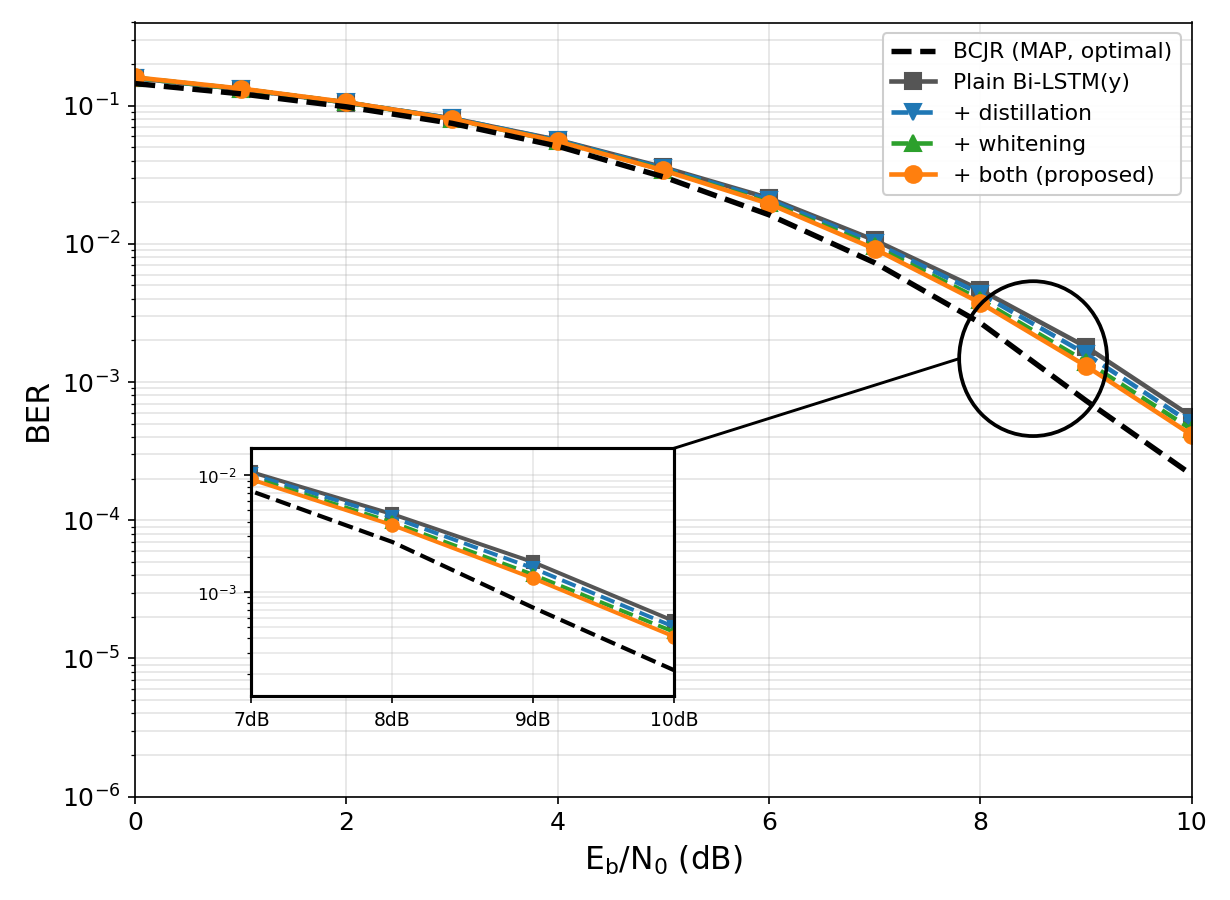}
\caption{BER vs.\ SNR at $\tau=0.6$ ($\beta=0.35$, $\kappa=4.7\times10^3$; strong
ISI): $1.73\times\!\to\!1.40\times$ MAP at $8$\,dB. Inset: $7$--$10$\,dB detail.}
\label{fig:ber06}
\end{figure}

\begin{figure}[t]
\centering
\includegraphics[width=0.97\columnwidth]{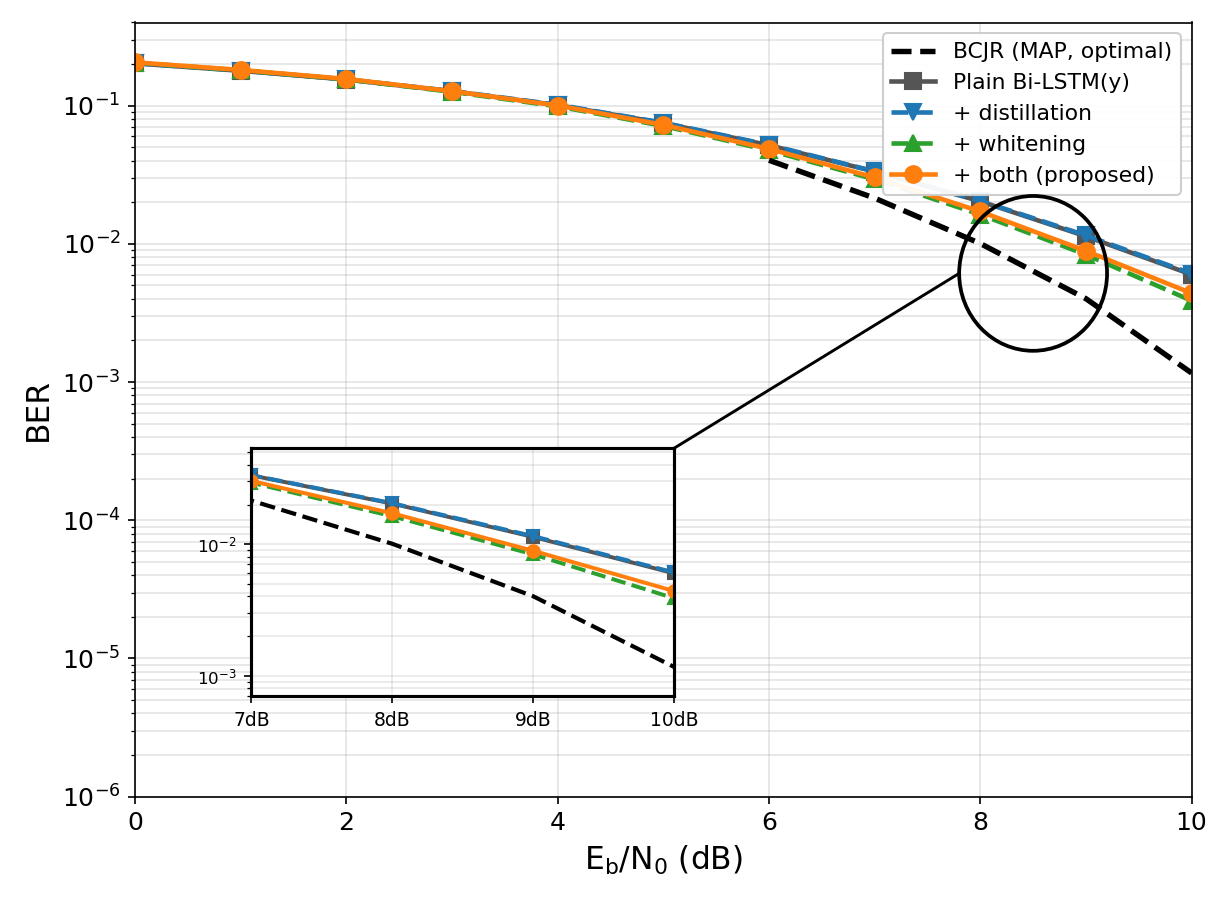}
\caption{BER vs.\ SNR at $\tau=0.5$ ($\beta=0.35$, $\kappa=7.7\times10^5$; severe
ISI, ill-conditioned whitening): $2.02\times\!\to\!1.71\times$ MAP at $8$\,dB.
Inset: $7$--$10$\,dB detail.}
\label{fig:ber05}
\end{figure}

Table~\ref{tab:main} and Figs.~\ref{fig:ber09}--\ref{fig:ber05} give the BER normalized by MAP at
$8$\,dB. The proposed method (whitening $+$ distillation) improves the baseline
by $23.7\%$ at $\tau=0.8$ (Fig.~\ref{fig:ber08}) and $19.2\%$ at $\tau=0.7$
(Fig.~\ref{fig:ber07}), reaching $1.05\times$ and $1.89\times$ MAP at $N=6$; increasing only the whitened window to $N=8$ reaches
$1.07\times$ and $1.47\times$. Across the $5$ seeds a paired $t$-test gives
$p=2.9\times10^{-5}$ ($\tau=0.8$) and $p=2.6\times10^{-3}$ ($\tau=0.7$); we
regard the block-wise and symbol-level tests of Section~\ref{subsec:robust} as
the primary evidence, since they do not rely on the seed being the unit of
analysis. Ablation shows the two components are individually weak
($-13.2\%$ for distillation and $-10.8\%$ for whitening at $\tau=0.8$) but
jointly strong ($-23.7\%$), consistent with the ceiling/reach picture. The same
ordering holds under stronger ISI at $\tau=0.6$ (Fig.~\ref{fig:ber06}), while at
$\tau=0.5$ (Fig.~\ref{fig:ber05}) the ridge-regularized whitening is already so
inexact that adding the teacher no longer helps.

\begin{table}[t]
\caption{Main results: BER normalized by MAP at $8$\,dB ($5$ seeds, $N=6$).
$\kappa$ is the condition number of the ISI matrix.}
\label{tab:main}
\centering
\begin{tabular}{lccccc}
\toprule
Model & $\tau{=}0.9$ & $\tau{=}0.8$ & $\tau{=}0.7$ & $\tau{=}0.6$ & $\tau{=}0.5$\\
$\kappa$ & $1.8$ & $6.3$ & $81$ & $4.7{\times}10^3$ & $7.7{\times}10^5$\\
\midrule
Plain $y$ (baseline)        & 1.04 & 1.37 & 2.34 & 1.73 & 2.02\\
$+$ distillation only       & 0.97 & 1.19 & 2.23 & 1.63 & 2.04\\
$+$ whitening only          & 1.01 & 1.23 & 2.00 & 1.47 & 1.62\\
$+$ both (proposed)         & 0.99 & \textbf{1.05} & 1.89 & \textbf{1.40} & 1.71\\
$+$ both, $N{=}8$           & --   & \textbf{1.07} & \textbf{1.47} & -- & --\\
\midrule
gain vs.\ baseline (\%)     & $-4$ & $-24$ & $-19$ & $-19$ & $-16$\\
\bottomrule
\end{tabular}
\end{table}

\subsection{Mechanism: the gain tracks conditioning}
\begin{figure}[t]
\centering
\includegraphics[width=0.99\columnwidth]{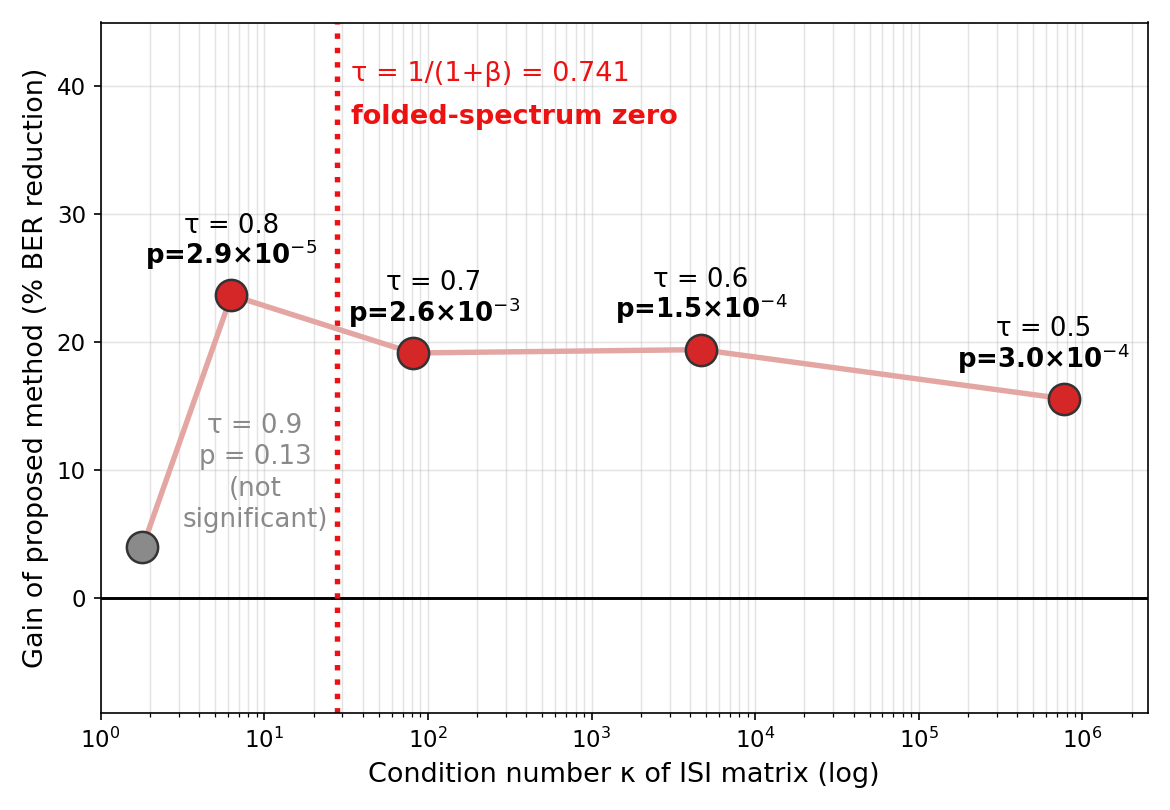}
\caption{Ill-conditioning is necessary but not sufficient: the gain is zero at
the well-conditioned $\tau=0.9$, peaks at $\tau=0.8$ ($\kappa=6.3$) and then
declines as $\kappa$ grows further. Dashed line marks the folded-spectrum zero
at $\tau=1/(1{+}\beta)$.}
\label{fig:cond}
\end{figure}

The folded spectrum of the RRC autocorrelation develops a zero exactly at
$\tau=1/(1{+}\beta)=0.741$; below it the ISI matrix becomes ill-conditioned
($\kappa$ in Table~\ref{tab:main}). Ill-conditioning is \emph{necessary} for the
gain but does not predict its size monotonically: the improvement is
$4.0\%,23.7\%,19.2\%,19.4\%,15.6\%$ for $\tau=0.9,\dots,0.5$, i.e.\ it is
essentially absent at the well-conditioned $\tau=0.9$ ($\kappa=1.8$), peaks
already at $\tau=0.8$ ($\kappa=6.3$) and then slowly declines even as $\kappa$
grows by five orders of magnitude (Fig.~\ref{fig:cond}). Two effects oppose each
other: worse conditioning means more headroom, but it also makes the whitening
transform itself harder to realize (the ridge $\delta$ grows to $10\%$ of $x_0$
at $\tau=0.5$). Notably the largest gain occurs on the \emph{well-conditioned}
side of the folded-spectrum zero. The absence of a gain where the model is
already well-conditioned is a \emph{sanity check} that the mechanism, not
overfitting, drives the improvement.

\subsection{Statistical robustness}\label{subsec:robust}
\begin{figure}[t]
\centering
\includegraphics[width=0.99\columnwidth]{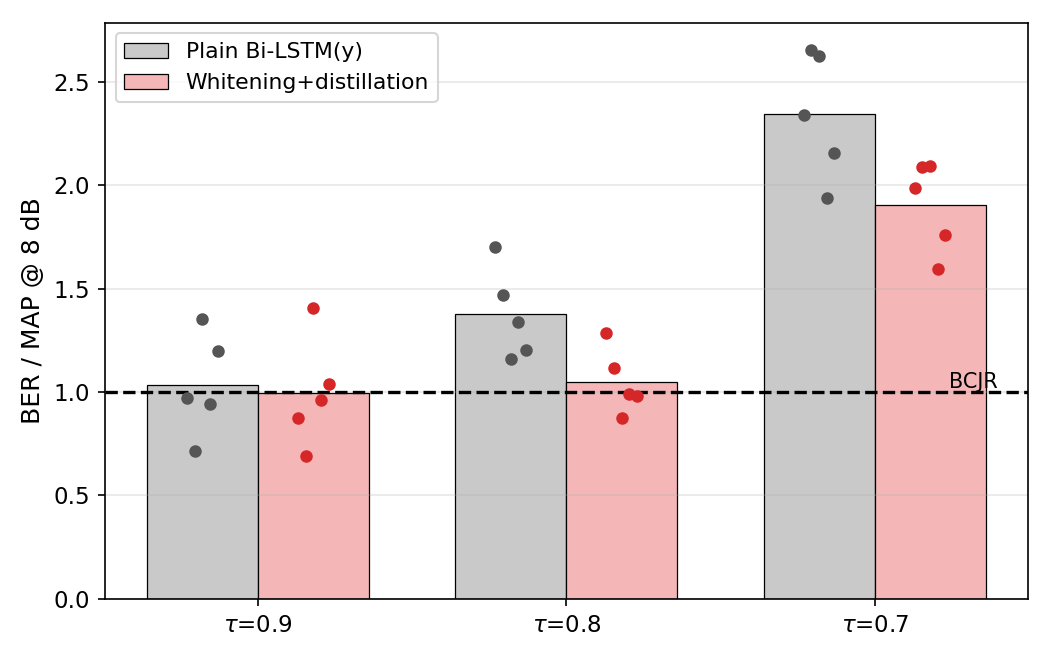}
\caption{BER normalized by MAP at $8$\,dB over five disjoint noise blocks (dots),
with the block mean (bars). The proposed detector (whitening $+$ distillation)
stays below the plain baseline at $\tau=0.8$ and $0.7$.}
\label{fig:blocks}

\vspace{2pt}
\includegraphics[width=0.99\columnwidth]{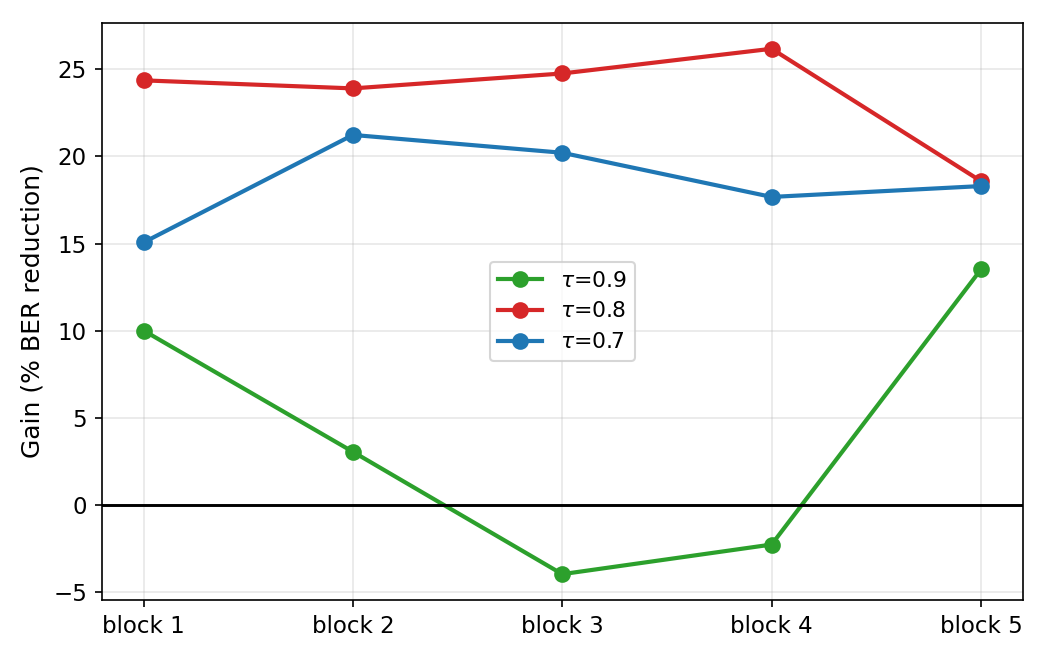}
\caption{Per-block gain (\% BER reduction) of the proposed detector. The gain is
positive in all five blocks at $\tau=0.8$ and $0.7$ ($5/5$), and mixed at the
well-conditioned $\tau=0.9$.}
\label{fig:perblock}

\vspace{2pt}
\includegraphics[width=0.99\columnwidth]{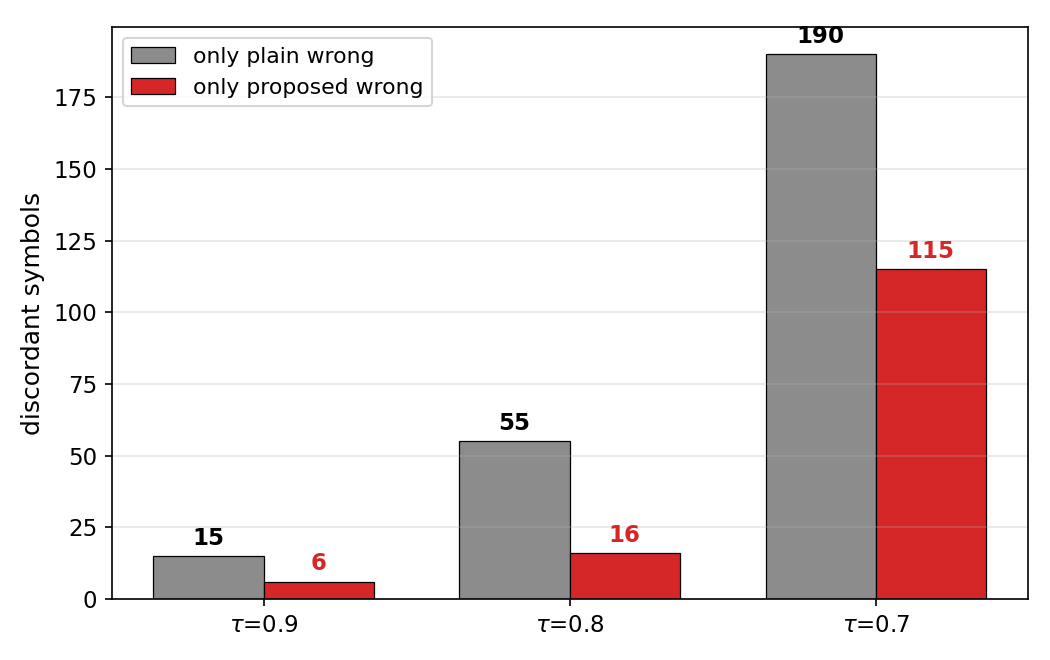}
\caption{Symbol-level McNemar counts on the common test set: symbols only the
plain model gets wrong vs.\ only the proposed model gets wrong. The imbalance is
significant at $\tau=0.8$ ($p=6.5\times10^{-6}$) and $0.7$ ($p=2.3\times10^{-5}$).}
\label{fig:mcnemar}

\vspace{2pt}
\end{figure}

Because the noise is i.i.d.\ in $k$, we split the held-out half into five
disjoint blocks (independent noise realizations) and additionally run a
symbol-level McNemar test (same test symbols, common-mode noise cancels). At
$\tau=0.8$ and $0.7$ the proposed method wins in $5/5$ independent blocks
(Figs.~\ref{fig:blocks} and~\ref{fig:perblock}) and the symbol-level McNemar test
is significant ($p=6.5\times10^{-6}$ and $2.3\times10^{-5}$; Fig.~\ref{fig:mcnemar});
at $\tau=0.9$ the difference is insignificant, as expected.

Two limitations bound these results. First, the whitening transform is realized
by a ridge-regularized Cholesky factorization whose regularizer grows to 10\% of
the main ISI tap at a compression factor of 0.5; past that point the transform
itself, rather than the network, limits what can be reached, which is why the
teacher stops helping there. Second, the block-wise and symbol-level robustness
tests were run only for compression factors 0.9, 0.8 and 0.7, so the gains
reported at 0.6 and 0.5 rest on the seed-level test alone. A numerically exact
factorization and the same robustness protocol at the lower compression factors
are therefore the natural next steps.

\section{Conclusion}\label{sec:conc}
For FTN detection, transferring the ISI-aware multi-window structure that works
for CNNs to a Bi-LSTM does not improve BER: the preprocessing adds no
information, extra branches only add bottlenecks, and a distillation diagnostic
shows the network is already window-optimal. The real limitation is the
observation model---colored matched-filter noise that violates the network's
conditional-independence assumption, worsening exactly where the ISI matrix is
ill-conditioned. Pre-whitening plus BCJR-posterior distillation, with the
architecture unchanged and only $+3.4\%$ parameters, closes most of the gap
($1.37\times\!\to\!1.05\times$ MAP at $\tau=0.8$ and
$2.34\times\!\to\!1.89\times$ at $\tau=0.7$ with $N=6$, the latter improving
to $1.47\times$ when only the whitened window is widened to $N=8$), validated
across independent noise realizations. The lesson is
that for recurrent FTN detectors the input representation and training target
matter more than architectural elaboration.

\section*{Acknowledgment}
This work was supported by The Scientific and Technological Research
Council of Türkiye (TÜBİTAK) under Project No.~122E236.

\end{document}